\documentclass[]{aa}  
\usepackage{graphicx,natbib}
\usepackage{txfonts}
\usepackage{epstopdf}

\newcommand{\ic}{{\tt iCosmo}}

\begin{document}
   \title{iCosmo: an interactive cosmology package}
\titlerunning{iCosmo, an Interactive Cosmology Package}
\authorrunning{Refregier, Amara, Kitching, Rassat}
   \author{A. Refregier\inst{1}, A. Amara\inst{2}, 
     T. D. Kitching\inst{3,}\inst{4}, A. Rassat\inst{1}}

   \institute{UMR 7158 (CNRS) / Service d'Astrophysique, AIM, CEA Saclay, 91191 Gif sur Yvette Cedex, France.
   \and
   Institute for Atronomy, ETH Hoenggerberg Campus, Physics Department, CH-8093 Zurich, Switzerland.
   \and
   Oxford Astrophysics, Department of Physics, Keble Road, Oxford, OX1 3RH, United Kingdom
   \and
   SUPA, University of Edinburgh, Royal Observatory Edinburgh, EH9 3HJ, United Kingdom
  } 
  
   \date{Draft: \today}

% \abstract{}{}{}{}{} 
% 5 {} token are mandatory
 
  \abstract
   {}
   {The interactive software package {\ic}, designed to perform cosmological
   calculations is described.}
   { {\ic} is a software package to perfom interactive cosmological
   calculations for the low-redshift universe. Computing distance 
   measures, the matter power spectrum, and the
   growth factor is supported for any values of the cosmological parameters.
   It also computes derived observed quantities for several cosmological probes
such as cosmic shear, baryon acoustic oscillations, and type Ia supernovae.
The associated errors for these observable quantities can be derived for customised surveys,
or for pre-set values corresponding to current or planned instruments.
The code also allows for calculation of cosmological forecasts with 
Fisher matrices, which can be manipulated to combine 
different surveys and cosmological probes. The code is written in the IDL language and thus
benefits from the convenient interactive features and scientific libraries available in this language. 
{\ic} can also be used as an engine to perform cosmological calculations in batch mode, and forms a convenient adaptive platform for the development of further cosmological modules. With its extensive documentation,
it may also serve as a useful resource for teaching and for newcomers to the field of cosmology. }
   {The {{\ic}} package is described with a number of examples and command sequences.
   The code is freely available with documentation at {\tt http://www.icosmo.org}, along with an
   interactive web interface
   and is part of the {\it Initiative for Cosmology}, a common archive for cosmological resources.}
   {}

   \keywords{cosmology --
             observables --
             numerical methods}

   \maketitle
%
%________________________________________________________________

\section{Introduction}

Cosmology is a rapidly advancing field thanks to recent progress in instrumentation,
numerical simulations, and theoretical methods. The last few decades have provided a previously innaccessible wealth of data with the advent of large surveys of the cosmic microwave background, galaxy clustering,
cosmic shear, clusters, and supernovae \citep[see][for more recent results and references therein]{wmap3:2006,wmap5:2008}.  One of the challenges of cosmology is to link the derived observable quantities from these cosmological surveys with the parameters of an underlying cosmological model.
%[\cite{yorke:1980}]. 
Another necessary task is to predict the performance of future surveys and future observational probes in order to maximise the potential of future experiments.  Existing publicly available computational tools \citep{CMBFAST,CAMB, COSMOMC} are widely used within the cosmological community, yet there is at present no a unified tool that considers the low-redshift universe and its associated probes.

In this paper, we present \ic, a software package that performs interactive
cosmological calculations. The code can also be used as an engine to perform
predictions for any cosmological model, and it also forms a convenient platform
for developing further cosmological modules. With its extensive documentation,
it may also serve as a useful resource for teaching and for newcomers in the field of cosmology.

The code calculates distance measures, the linear and non-linear matter power
spectrum, the growth factor, volume elements and other quantities for any CDM cosmological model.
It also computes derived observable quantities for several cosmological probes
such as cosmic shear, baryonic acoustic oscillations (BAO), and type Ia supernovae.
The associated errors for these observables can be described for arbitrary survey parameters,
or alternatively, for a number of preset values corresponding to planned surveys and experiments.
The code also provides forecasts for future cosmological surveys, by allowing the user to compute, manipulate, and combine Fisher matrices. The code is written in the IDL language, so it benefits from the simple syntax, large scientific libraries and convenient interactive plotting environment of this language. In particular, all the above calculations can be completed in only a few lines of code.

The code is freely available at {\tt http://www.icosmo.org}
with full documentation and help files. It is part of the {\it Initiative for Cosmology} web site, %a global archive for cosmological resources. An 
and it includes an interactive web interface %to \ic is also available at this website 
\citep[see description in][]{Kitching:2008}.

In this paper, we describe the main features of \ic~  and provide several examples
illustrating its use. In section~\ref{general}, we describe the main conventions and architecture
of the code. In section~\ref{evol} we describe how the code can quickly calculate the evolution of
the main cosmological quantities, such as distance measures and growth factor, as a function
of redshift. In section~\ref{probes}, we describe how \ic~ calculates the derived observable quantities
for cosmic shear, BAO and type Ia supernovae. Section~\ref{fisher}
shows how the code can compute and manipulate Fisher matrices. Our conclusions 
and possible future developments are presented in section~\ref{conclusion}.

%\section{Modern Cosmological Tools}
%\ubsection{Cosmological Probes} \label{sec:cosmology}

\section{General presentation}
\label{general}

The \ic~ code is a package of routines written in the {\tt IDL} language. It is
divided into several directories corresponding to the different levels of the
calculations:
\begin{itemize}
\item {\tt general, plotting:} General utility and plotting routines;
\item {\tt cosmo, expt}: Routines to define fiducial cosmological and survey parameters and compute basic cosmological quantities;
\item {\tt lensing}, {\tt bao}, {\tt sne}: Routines to compute cosmic shear, BAO and type Ia supernovae observables;
\item {\tt fisher}: Routines to create, plot and manipulate Fisher matrices for cosmological forecasts.
\end{itemize}

\begin{table*}[htdp]
\caption{Main \ic routines and a brief description of their use.}
\label{tab:routines}
\begin{center}
\begin{tabular}{lll}
\hline
\hline
Directory & Routine & Description \\
\hline
{\tt cosmo} & {\tt set\_fiducial} & Create fiducial parameter structure \\ 
                  & {\tt mk\_cosmo} & Compute basic cosmological quantities \\
              	& {\tt get\_pk} & Extracts 3D power spectrum \\
{\tt expt} &{\tt mk\_survey} & Compute survey parameters\\
{\tt lensing} & {\tt mk\_cl\_tomo} & Compute shear power spectrum \\
 & {\tt mk\_cl\_cov\_tomo} & Compute covariance errors\\
 & {\tt mk\_fisher\_lens} & Compute cosmic shear Fisher matrix and errors\\
{\tt bao} & {\tt mk\_bao} & Compute BAO distance measures\\
 & {\tt mk\_bao\_cov} & Compute BAO covariance errors\\
 & {\tt mk\_fisher\_bao} & Compute BAO Fisher matrix and errors\\
{\tt sne} & {\tt mk\_sne} & Compute SNe magnitude-redshift relation \\
 & {\tt mk\_sne\_cov} & Compute SNe covariance errors\\
 & {\tt mk\_fisher\_sne} & Compute SNe Fisher matrix\\
{\tt fisher} & {\tt check\_matrix} & Check matrix is positive definite\\
 & {\tt comb\_fisher} & Combine two Fisher matrices\\
 & {\tt margin\_fisher} & Marginalise over unwanted parameters\\
{\tt plotting} & {\tt plt\_cosmo}& Plot cosmology parameters\\
 & {\tt plt\_pk}& Plot 3D matter power spectrum \\
 & {\tt plt\_sv}& Plot survey properties\\
 & {\tt plt\_cl}& Plot cosmic shear correlation function\\
 & {\tt plt\_bao}& Plot BAO distance measures\\
& {\tt plt\_sne}& Plot supernova magnitude-redshift relation\\
& {\tt plt\_fisher}& Plot Fisher matrix errors (combined 1D and 2D errors)\\
& {\tt plt\_fisher\_1p}& Plot 1D likelihood errors\\
& {\tt plt\_fisher\_2p}& Plot 2D error ellipses \\
 
\hline
\end{tabular}
\end{center}
\label{default}
\end{table*}%

The routines in each of the directories make use of the variable structures in IDL. Their functionality
is indicated by the first few letters of their name with the following conventions:
\begin{itemize}
\item {\tt set}: set up basic structures with calculation variables,
\item {\tt mk}: make or compute a new structure using the information in an input structure,
\item {\tt get}: get a substructure or derived quantity from an existing structure,
\item {\tt plt}: plot,
\item {\tt rd}, {\tt wt}: read and write data from and into a file.
\end{itemize}

A description of the main \ic~ routines and associated directories can be found in Table~\ref{tab:routines}.
Detailed explanations for any of the routines can be obtained
by typing {\tt icosmo\_help,} `{\it routine name}' at the IDL command line.
A {\tt readme} file is available in the distribution and gives installation instructions and a
quick start tutorial.

Table~\ref{tab:sequence} shows an example of a typical call sequence for \ic, with the different levels
of the code separated by horizontal lines. The following sections describe
the use of each level of the code with examples of call sequences and output figures. 

\begin{table*}[htdp]
\caption{Example of a typical call sequence for \ic, with different levels of the code separated
by horizontal lines.}
\label{tab:sequence}
\begin{center}
\begin{tabular}{cll}
\hline \hline
& Instruction & Description\\
\hline
1& {\tt fid=set\_fiducial(cosmo\_in=\{omega\_m:0.31\})} & Set fiducial parameters, choosing a non-default value of the parameter $\Omega_m$\\
2 & {\tt cosmo=mk\_cosmo(fid)} & Compute basic cosmological quantities (e.g., distance measures, power spectrum) \\ 
3 & {\tt plt\_cosmo,cosmo,`z',`da'} & Plot angular-diameter distance $D_{A}(z)$ as a function of redshift $z$\\
4 & {\tt plt\_pk,cosmo,z=0} & Plot non-linear matter power spectrum $P(k)$ at $z=0$\\
\hline
5 & {\tt sv=mk\_survey(fid,'sv1')} & Read survey parameters for the preset DUNE experiment\\  
6 & {\tt  cl=mk\_cl\_tomo(fid,cosmo,sv)} & Compute cosmic shear power spectrum $C_{l}$\\
7 & {\tt cl\_cov=mk\_cl\_cov\_tomo(fid,cl,sv)} & Compute cosmic shear power spectrum errors\\
8 & {\tt plt\_cl,cl,[0,0],cl\_cov,/errors} & Plot shear power spectrum and associated errors\\
\hline
9 & {\tt fish=mk\_fisher\_lens(fid,sv)} & Compute Fisher matrix for the specified lensing survey\\ 
10 & {\tt margin\_fisher,fish,fish2,[0,1,1,0,0,0,0,0]} & Marginalise Fisher matrix over unwanted parameters\\
11 & {\tt  plt\_fisher\_2p,fisher2} & Plot Fisher matrix error contraints for interesting parameters $w_0$ and $w_a$\\
\hline
\end{tabular}
\end{center}
\label{default}
\end{table*}%

\section{Basic cosmological quantities}
\label{evol}
The first level in \ic~ defines the cosmological model and computes basic cosmological quantities; see, e.g., \cite{Peacock:1999}
for their definitions and \cite{Refregier:2004} for conventions. The first step is to define the
fiducial parameter structure using the {\tt set\_fiducial} function (see Inst.~1 in Table~\ref{tab:sequence}).
The resulting {\tt fid} structure contains the following three substructures:
\begin{itemize}
\item {\tt cosmo}: input cosmological parameters,
\item {\tt expt}: experiment parameters, such as survey area, number of galaxies or SNe as a function of redshift,
or preset surveys,
\item {\tt calc}: calculation parameters, such as redshift and $k$-range, fitting functions, etc.
\end{itemize}
These parameters have default values that can be easily modified using the syntax of
instruction~1 in Table~\ref{tab:sequence}.

The next step is to compute all the basic cosmological quantities using the
{\tt mk\_cosmo} function (see Inst.~2 in Table~\ref{tab:sequence}). The
resulting {\tt cosmo} structure is organised into several substructures:
\begin{itemize}
\item {\tt const}: all original fiducial quantities, as well as constant quantities derived from the fiducial parameters
(e.g., the Hubble radius),
\item {\tt evol}: Evolving scalar quantities such as distance measures, the growth factor, tabulated
as a function of redshift,
\item {\tt pk}: Linear and non-linear matter power spectrum $P(k,z)$ tabulated as a function of 
wavenumber $k$ and redshift $z$.
\end{itemize}

\begin{figure}[htbp]
\begin{center}
\includegraphics[width=7.5cm,angle=0]{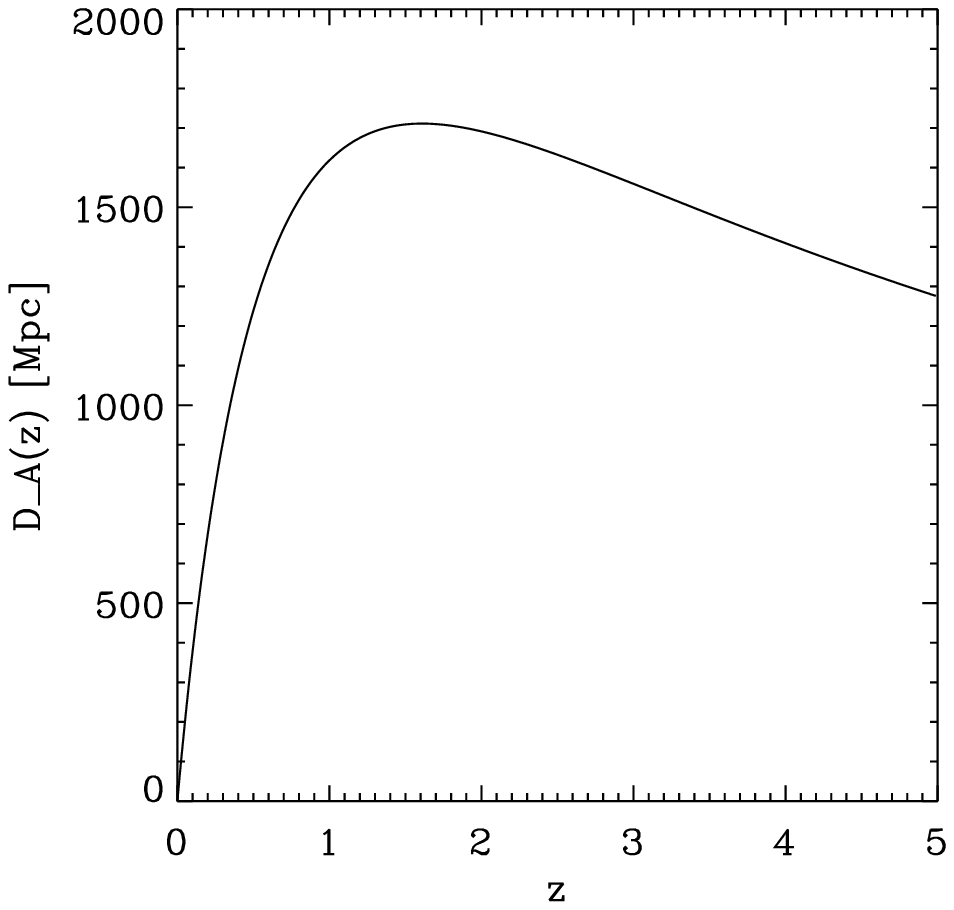}
\caption{Example of an \ic~ output showing the angular-diameter distance
$D_A(z)$ as a function of redshift. See text for command sequence. }
\label{fig:da}
\end{center}
\end{figure}

The evolving scalar quantities can then be plotted easily using the {\tt plt\_cosmo} routine
(see Inst.~3 in Table~\ref{tab:sequence}). As an example, the following
call sequence: \\
\noindent {\tt $\rangle$ fid=set\_fiducial(cosmo=\{w0:-0.9\},calc=\{fit\_tk:1\})}\\
{\tt $\rangle$ cosmo=mk\_cosmo(fid)}\\
{\tt $\rangle$ plt\_cosmo,cosmo,`z',`da'}\\
produces figure~\ref{fig:da} which shows the angular diameter distance $D_A(z)$
as a function of redshift $z$, defined in a flat universe by 
\begin{equation} D_A(z) = \frac{\chi(z)}{1+z},~~~~~~~\chi(z) = \int_0^z \frac{c{\rm d}z'}{H(z')},\end{equation}
where $\chi(z)$ represents the comoving radial distance, $H(z)$ the Hubble parameter, and $c$ the speed of light.
In all figures below, the fiducial cosmology is
set to [$h = 0.7, \Omega_b = 0.045, \Omega_m =0.3, \Omega_\Lambda=0.7, w_0=-0.95, w_a=0.0, n=1.0, \tau=0.09, \sigma_8=0.8$], unless otherwise stipulated.  For example, the call sequence above changes $w_0$ to $-0.9$, which is used in the rest of this paper.

\begin{figure}[htbp]
\begin{center}
\includegraphics[width=8.cm,angle=0]{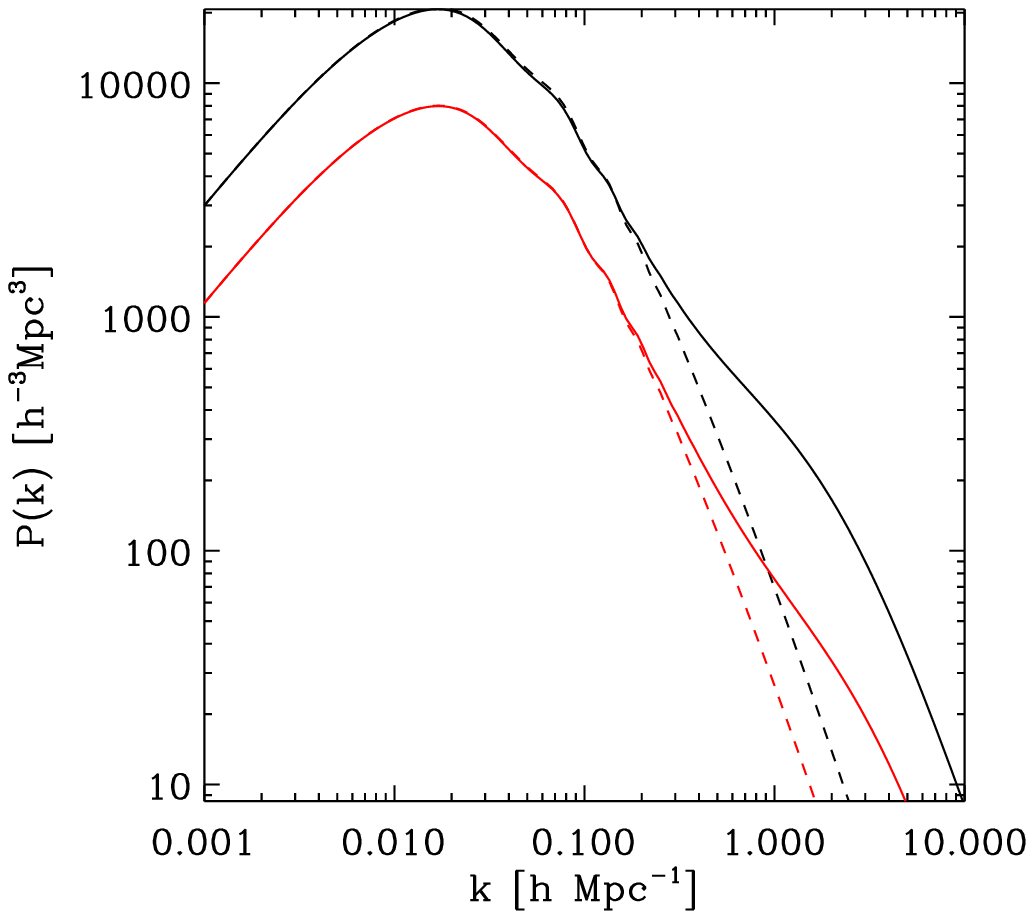}
\caption{Matter power spectrum at $z=0$ and $1$, for top and bottom lines,
respectively. Linear and non-linear power spectra are shown as dashed
and solid lines. These plots were derived using the {\tt plt\_pk} routine.}
\label{fig:pk}
\end{center}
\end{figure}

As an example of the plotting functionality, the plotting routine {\tt plt\_pk} can be used to plot the matter power spectrum $P(k,z)$ as a
function of wavenumber $k$ at a redshift $z$, where $P(k)$ is defined as \begin{equation} \left< \delta({\bf k})\delta^*({\bf k'})\right>=(2\pi)^3 \delta^3\left({\bf k}-{\bf k'}\right)P(k),\end{equation} and $\delta({\bf k})$ represents matter fluctuations in Fourier space.  For example, figure \ref{fig:pk} 
shows the linear and non-linear power spectrum at $z=0$ and $1$ and is produced
by the additional sequence:\\
\noindent {\tt $\rangle$ plt\_pk,cosmo,z=0,xran=[0.001,10.] }\\
{\tt $\rangle$ plt\_pk,cosmo,z=0,/over,/linear,linestyle=2}\\
{\tt $\rangle$ plt\_pk,cosmo,z=1,/over,color=2}\\
{\tt $\rangle$ plt\_pk,cosmo,z=1,/over,/linear,linestyle=2,color=2}\\
The keyword {\tt over} can be used to overlay several results
on a single figure.

\section{Derived cosmological observable quantities}
\label{probes}
The next level of \ic~ consists in the calculations of the derived observables related
to different cosmological probes. At present, cosmic shear, baryon acoustic 
oscillations, and type Ia supernovae are supported.  The underlying physics of each of these probes is described below. 

\subsection{Cosmic Shear}

As light travels towards us from distant galaxies, its path is perturbed by intervening matter along the line of sight. This causes distortions in the background images and is known as gravitational lensing. To first order this can be described by a distortion matrix: 
\begin{equation} \Psi_{ij} \equiv \frac{\partial (\delta \theta_i)}{\partial \theta_j}\equiv \left( \begin{array}{cc}
\kappa +\gamma_1 & \gamma_2 \\
\gamma_2  & \kappa -\gamma_1\end{array} \right), \end{equation} where $\delta \theta_i(\theta)$ is the deflection vector that results from the lensing.  The distortion matrix captures two effects of  gravitational lensing: the observed image can be dilated or contracted, which is described by the convergence $\kappa$, and the image can also be stretched and compressed along (and at $45^\circ$ from) the $x$-axis, which is described by the shear component $\gamma_1$ ($\gamma_2$).

Cosmic shear - the statistical measure of shear ($\gamma_1$, $\gamma_2$), uses the two point function of the shear field as a cosmological observable, given by 
\begin{equation}
C_{ij}(\ell) = \frac{9}{16}\left(\frac{H_0}{c}\right)^4\Omega_m^2\int_0^{\chi_h}d\chi\left[\frac{g_i(\chi)}{ar(\chi)}\right]\left[\frac{g_j(\chi)}{ar(\chi)}\right]P\left(\frac{l}{r},\chi\right) ,
\label{eq:shear}\end{equation}
where the weighting function is given by 
\begin{equation} g_i(\chi)=\int_\chi^{\chi_h}d\chi' n_i(\chi') \frac{r(\chi) r(\chi' -\chi)}{r(\chi'}, \end{equation}
where $r=\frac{D_A}{a}$ and $D_A$ is the angular-diameter distance. In Equation \ref{eq:shear}, the auto-correlation is given for $i=j$ and the cross-correlation for $i\ne j$, where $i$ and $j$ correspond to galaxy populations at different redshifts.  The galaxy distribution function is given by $n_i(\chi)$, i.e. the probability of finding a galaxy at a distance $\chi$, and is normalised to $\int d\chi n_i(\chi)=1$.

%The shear signal arising from different redshift will be correlated with each other.  This correlation contains useful signal and can be quantified by the cross-correlation signal: 
%\begin{equation}
%C_{ij}(\ell) = \frac{9}{16}\left(\frac{H_0}{c}\right)^4\Omega_m^2\int_0^{\chi_h}d\chi\left[\frac{g_i(\chi)}{ar(\chi)}\right]\left[\frac{g_j(\chi)}{ar(\chi)}\right]P\left(\frac{l}{r},\chi\right) ,
%\end{equation}

In \ic, cosmic shear is implemented using shear power spectrum tomography as described in  
Refs.~\cite{Hu:1999}, \cite{Hu:2004}, \cite{Refregier:2004} and \cite{Amara:2007}.  The uncertainties $\Delta C_{ij}(\ell)$ are calculated under the assumption that the two-point function is Gaussian.

\subsection{Baryon Acoustic Oscillations}
In the early Universe, just before decoupling, baryons and photons were coupled through Thomson scattering and Coulomb interactions.  The photon-baryon fluid was subject to two competing effects: collapse due to gravitational instability and repulsion due to outward pressure.  These two effects created oscillations in the fluctuation field, visible today in the Cosmic Microwave Background \citep[CMB,][]{Peebles:1970,Sunyaev:1970}, as well as in the galaxy distribution.  In Fourier space, this effect corresponds to a series of acoustic peaks, dubbed Baryon Acoustic Oscillations (BAOs). They can be visualised by considering the ratio of a matter power spectrum with baryons, to one without: $P^{\rm b}(k)/P^{\rm no b}(k)$, and in this case are also called `wiggles'.% (see Figure \ref{fig:bao} of ratio).

By measuring the BAOs in the matter power spectrum, it is possible to extract a characteristic BAO scale present in both the radial ($y$) and tangential ($y'$) directions \citep[see section 2.2 of][]{Blake:2006}:
\begin{equation} ~~~~~~~y=\frac{r(z)}{s},~~~~~~y'=\frac{c}{H(z)s} ,\end{equation}
where $r(z)$ corresponds to the comoving radial distance, $H(z)$ to the Hubble parameter, $s$ to the comoving sound horizon at recombination and $c$ to the speed of light. The reviews by \cite{WGFC} and \cite{DETF} concluded that the measurement of the BAO scale was a fundamental tool for future precision cosmology.

To use the information contained in the BAO scale, it is necessary to have an estimate of the accuracy with which a given survey can measure the scale.  In the current version of \ic, we use the analytic expression derived in \cite{Blake:2006} for the accuracy of the measurement on the BAO scale \citep[though there are several approaches for using information from BAOs, as shown in][]{Rassat:2008bao}.  This analytic expression was derived from simulations and depends on the central redshift $z$ of the survey, the total volume $V$ of the survey, and the average number density of galaxies.  For the moment, photometric redshift errors are not included in the \ic~ BAO module, though it is given in the accuracy expression of \cite{Blake:2006}.  The expression for the accuracy measurement is given by equations 6-9 in \cite{Blake:2006}.  
%\begin{equation}\end{equation}
\subsection{Type Ia Supernovae}

%Type Ia supernovae arise from material accretion on white dwarves by a binary companion; the supernova explosion happens as the white dwarves reaches the Chandrasekhar limit.  Since it is a standard mass limit that causes their explosion, 
Type Ia supernovae are assumed in cosmology to behave as \emph{standard candles}; i.e., the rate with which the brightness of a supernova declines is assumed proportional to its intrinsic luminosity.  Thus by measuring the light curves of distant supernovae, it is possible to measure their distance, and by obtaining spectra one can compare their distance-redshift relation, which is highly dependent on cosmological parameters.

The derived observable for type Ia supernovae is therefore the apparent magnitude $m(z)$ at a given redshift $z$, related to the luminosity distance by 
\begin{equation} m(z) = \mathcal{M}+5{\rm log}_{10}D_L(z), \end{equation}% Studies in the late '90s \cite{Riess:1998,Perlmutter:1998,Schmidt:1998} made the surprising conclusion that the Universe might be accelerating.
where the $H_0$-independent luminosity distance $D_L(z)$ is given by \begin{equation}D_L(z)\equiv (H_0/c)(1+z)\chi(z).\end{equation} The normalisation parameter $\mathcal{M}$ is given by \begin{equation}\mathcal{M}= M-5{\rm log}_{10}(H_0/c)+{\rm cst},\end{equation}where $M$ is the absolute magnitude.

The uncertainty on the magnitude is given by \cite{Kim:2004}:
\begin{equation} \Delta m(z) = \sqrt{\sigma_m^2 + \left(\frac{5\sigma_\nu}{cz\ln 10}\right)^2 + N_z \delta^2_m},\end{equation} where $\sigma_\nu$ accounts for the scatter in peculiar velocities, $\sigma_m$ the observed variance, and $N_z$ is the number of supernovae redshift bins

%For BAO, radial and tangential distance scales are calculated, along with  their errors derived from the Blake et al. [\cite{Blake:2006}] fitting formula (see section \ref{sec:cosmology:bao}). 
%We calculate magnitude-redshift relation for supernovae using the formalism outlined in Refs.~[\cite{Tegmark:1998,Huterer:2001}]and include both intrinsic scatter and observed variance.

\subsection{Derived cosmological observables with \ic}

A typical call sequence is given in Instructions~5-8 in Table~\ref{tab:sequence}. The sequence first defines
the survey using {\tt mk\_survey} and then computes the derived observable
using {\tt mk\_cl\_tomo}, {\tt mk\_bao}, {\tt mk\_sne}, for lensing tomography, BAO and SNe, respectively.
The statistical errors in these observables for the fiducial survey can be derived using
the correspondng {\tt mk\_cov} routines listed in Table~\ref{tab:routines}.

\begin{figure}[htbp]
\begin{center}
\includegraphics[width=8.cm,angle=0]{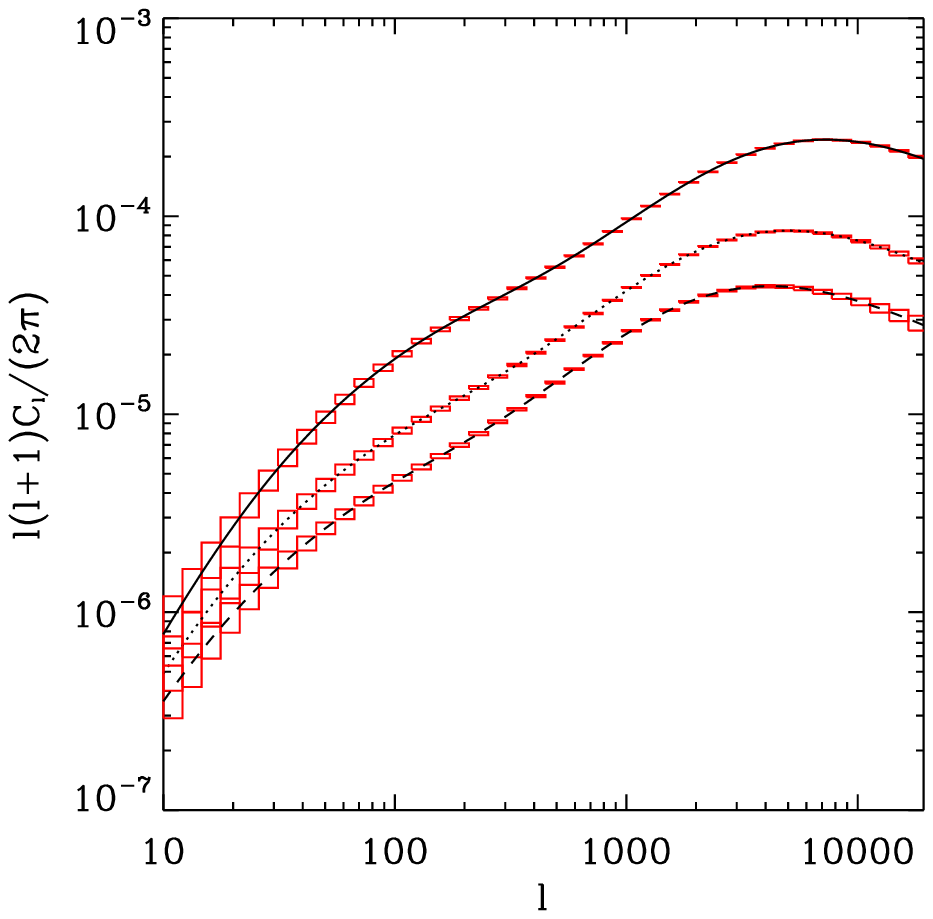}
\caption{The cosmic shear power spectrum and associated $1\sigma$ error
bars for the DUNE/Euclid cosmic shear survey for the $1^{\rm st}$ (upper line) 
and $2^{\rm nd}$ (middle line) 
tomographic bins (with median redshifts of 0.68 and 1.36 respectively) as well as the cross-correlation between the bins (lower line).}
\label{fig:wl}
\end{center}
\end{figure}

As an example, the following produces figure~\ref{fig:wl}, a plot of
cosmic shear power spectra for DUNE \citep{Refregier:2008} with its associated $1\sigma$ error bars:\\

\noindent {\tt $\rangle$ fid=set\_fiducial('DUNE', expt\_in=\{sv1\_n\_zbin:2\})}\\
{\tt $\rangle$ cosmo=mk\_cosmo(fid)}\\
{\tt $\rangle$ sv=mk\_survey(fid,'sv1')}\\
{\tt $\rangle$ cl=mk\_cl\_tomo(fid,cosmo,sv)}\\
{\tt $\rangle$ cl\_cov=mk\_cl\_cov\_tomo(fid,cl,sv)}\\
{\tt $\rangle$ plt\_cl,cl,[1,1],cl\_cov,/errors,yran=[1e-7,1e-3]}\\
{\tt $\rangle$ plt\_cl,cl,[0,1],cl\_cov,/errors,/over,linestyle=1}\\
{\tt $\rangle$ plt\_cl,cl,[0,0],cl\_cov,/errors,/over,linestyle=2}\\

The different power spectra correspond to two different tomographic
redshift bins with median redshifts 0.68 and 1.36, as well as their cross-power spectra. Similar call sequences can produce predictions and errors for the BAO distance measures and for SNe Hubble diagrams.

\section{Fisher Matrices}
\label{fisher}
The final level of \ic~ is related to the computation and  manipulation of Fisher matrices to 
assess the constraints on cosmological parameters that can be achieved with
future surveys.
\subsection{The Fisher Information Matrix}

It is possible to forecast the precision with which a given experiment can constrain cosmological parameters using the Fisher Information Matrix \citep[FIM, see][for a detailed derivation]{Tegmark:1997th}.  A method based on the FIM requires the following ingredients as input: \begin{itemize}
\item A set of cosmological parameters $\vec{\theta}=(\theta_1, \theta_2, ..., \theta_i)$ for which one requires predicted constraints;
\item A choice of underlying fiducial cosmology;
\item A set of measurements $\vec{x} = (x_1, x_2, ..., x_n)$, for e.g., this could be the shear power spectrum $C(\ell)$ over a range $\ell=1 ... n$; 
\item A cosmological model for how the observable depends on cosmological parameters, i.e. a model for calculating $C(\ell) = C(\ell, \vec{\theta})$;
\item An estimate of the uncertainty on the observable, i.e. in our example an estimate of $\Delta C(\ell)$.  This may depend on the experiment (instrument noise, shot noise, etc) as well as on the data estimator (cosmic variance).
\end{itemize}

With the above inputs, it is possible to calculate the FIM, whose components are denoted as $F_{ij}$: 
\begin{equation} 
F_{ij} = \left< \frac{\partial^2{\bf L}}{\partial \theta_i \partial \theta_j}\right>,
\end{equation} where ${\bf L} = -\ln L$, and $L=L(\vec{x}, \vec{\theta})$ is the likelihood function or the probability distribution of the data $\vec{x}$, which depends on some cosmological parameter set $\vec{\theta}$.

In practise, for a given observable $C(\ell, \vec{\theta})$, assumed to have Gaussian errors, the Fisher matrix can be calculated using 
\begin{equation} F_{ij} \simeq \sum_\ell \frac{1}{\left[\Delta C(\ell)\right]^2}\frac{\partial C(\ell, \vec{\theta})}{\partial \theta_i}\frac{\partial C(\ell,\vec{\theta})}{\partial \theta_j}.\end{equation}
For example, the coefficients of the Fisher matrix for the supernovae calculation is obtained by calculating \citep[see][]{Tegmark:1998,Huterer:2001}
\begin{equation} F^{\rm SN Ia}_{ij} = \sum_z^{N_z}\frac{1}{\left[\Delta m(z)\right]^2}\frac{\partial m(z)}{\partial \theta_i}\frac{\partial m(z)}{\partial \theta_j},\end{equation} where the sum is over redshift bin.

When a vector of parameters are allowed to vary, \emph{marginalised} parameter constraints $\Delta \theta^{\rm marg}_i$ can be obtained by 
\begin{equation} \Delta \theta^{\rm marg}_i \ge \sqrt{\left(F^{-1}\right)_{ii}}.\end{equation}
This equation provides a \emph{lower} limit on the constraints one can expect to attain for a given survey and fiducial cosmology.
When all other cosmological parameters are fixed, the constraint on a parameter $\theta^{\rm fix}_i$ can be estimated by 
\begin{equation}\Delta \theta^{\rm fix}_i \ge \frac{1}{\sqrt{F_{ii}}}. \end{equation}

\subsection{Calculating Fisher Matrices with \ic}

 For this purpose, Fisher matrices can first be computed using the {\tt mk\_fisher}
routines for each probe (see Table~\ref{tab:routines}). The computation is done
by taking excursions from the fiducial cosmological model for each of the parameters
of interest. Fisher matrices can be combined using the {\tt combine\_fisher} routine 
to derive joint cosmological constraints with several surveys and probes. The routine
{\tt margin\_fisher} can be used to marginalise over unwanted parameters. The resulting
contraints can be plotted using the {\tt plt\_fisher} routines.

A typical call sequence to compute and plot Fisher matrices can be found in instructions 9-11 in Table~\ref{tab:sequence}.  As an example, the following sequence computes the Fisher
matrices for a DUNE cosmic shear survey, a full sky BAO spectroscopic survey (20,000 deg$^2$),
and their combination:\\

\noindent {\tt $\rangle$ fid\_{lens}=set\_fiducial('DUNE')}\\
{\tt $\rangle$ fid\_{bao}=set\_fiducial('bao\_halfsky')}\\
{\tt $\rangle$}\\
{\tt $\rangle$ sv\_lens=mk\_survey(fid\_lens,'sv1')}\\
{\tt $\rangle$ sv\_bao=mk\_survey(fid\_bao,'sv2')}\\
{\tt $\rangle$}\\
{\tt $\rangle$ f\_lens=mk\_fisher\_lens(fid\_lens,sv\_lens)}\\
{\tt $\rangle$ f\_bao=mk\_fisher\_bao(fid\_bao,sv\_bao)}\\
{\tt $\rangle$ f\_comb=comb\_fisher(f\_lens,f\_bao)}\\

\begin{figure}[htbp]
\begin{center}
\includegraphics[width=8.cm,angle=0]{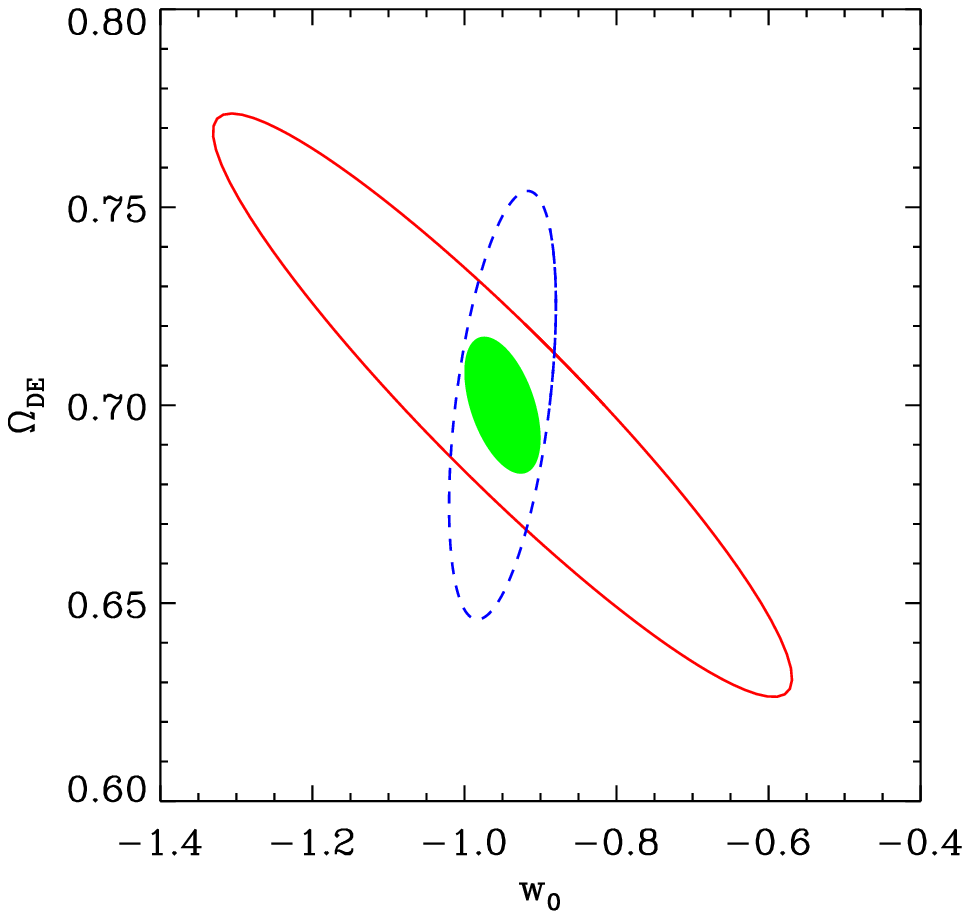}
%\vspace{8.cm}
\caption{Marginalised 68\%CL constraints on the dark energy parameters $\Omega_{DE}$ and $w_0$, expected
for the DUNE cosmic shear (blue), a full sky BAO survey (red) and their combination (solid green). This figure was derived using the Fisher matrix routines of \ic.}
\label{fig:fisher}
\end{center}
\end{figure}

The following call sequence produces figure~\ref{fig:fisher}
showing the 68\%CL constraints on the dark energy parameters $\Omega_{DE}$ and $w_0$
expected for each surveys separately (blue and red ellipse) and jointly (solid green ellipse).\\

\noindent {\tt $\rangle$ margin\_fisher,f\_lens,f\_lens2,[0,1,0,0,0,0,0,1]}\\
{\tt $\rangle$ margin\_fisher,f\_bao,f\_bao2,[0,1,0,0,0,1] }\\
{\tt $\rangle$ margin\_fisher,f\_comb,f\_comb2,[0,1,0,0,0,0,0,1]}\\
{\tt $\rangle$ plt\_fisher\_2p,f\_bao2,/nofill,color=2}\\
{\tt $\rangle$ plt\_fisher\_2p,f\_lens2,/nofill, \$ linestyle=2,/over}\\
{\tt $\rangle$ plt\_fisher\_2p,f\_comb2,/over,color=3}\\

The {\tt margin\_fisher} routine was used to retain only the parameters to be
plotted, and marginalise over all the other parameters.

\section{Conclusions}
\label{conclusion}
The \ic~ package provides a convenient and flexible interactive tool for 
cosmological calculations, and it can also be used as a computation engine
in batch mode. With all the built-in IDL libraries, it is also a convenient
platform for developing other cosmological routines and for teaching. The code, along
with an interactive web tool \cite{Kitching:2008} and various cosmological resources are freely available at
at  {\tt http://www.icosmo.org}.

In the future, we plan to add more features to the code, such as
the halo model, higher order clustering statistics, and interfaces with CMB
Boltzmann codes. Contributions from the community are encouraged.

%\appendix

\begin{acknowledgements}    
The authors thank the many people who have contributed to \ic~ over
the years, particularly, Richard Massey, David Bacon, Joel Berge, Florian Pacaud,
Romain Teyssier, Jean-Baptiste Juin, Dominique Ivon, Ivan Debono, Sarah Bridle, Andrew Hodgson, Benjamin Joachimi, Chaz Shapiro, Andy Taylor, Licia Verde, Martin Kunz, and Francisco Castander.  TDK is supported by STFC Rolling Grant Number RA0888.
\end{acknowledgements}

\bibliographystyle{aa}
\bibliography{references}
%\begin{thebibliography}{}
%
%\bibitem[1980a]{yorke80a} Yorke, H. W. 1980a,
%      A\&A, 86, 286
%
%\end{thebibliography}

\end{document}